\documentclass[pra,superscriptaddress,twocolumn,notitlepage]{revtex4-1}
\usepackage{amsmath}
\usepackage{amsfonts}
\usepackage{amssymb}
\usepackage{float}
\usepackage{mathrsfs}

\usepackage{dsfont}
\usepackage{graphicx}
\usepackage[usenames,dvipsnames]{xcolor}

\renewcommand{\d}{\dagger}
\newcommand{\dt}[1]{\ensuremath{\frac{\mathrm{d} #1}{\mathrm{d} t}}}

\newcommand{\ii}{\ensuremath{\mathrm{i}}}
\newcommand{\ee}{\ensuremath{\mathrm{e}}}
\newcommand{\bra}[1]{\ensuremath{\left\langle#1\right\rvert}}
\newcommand{\ket}[1]{\ensuremath{\left\lvert#1\right\rangle}}

\newcommand{\ketbra}[2]{\ensuremath{\left| #1\right\rangle\!\left\langle #2\right|} }

\newcommand{\avg}[1]{\left \langle {#1} \right\rangle}

\newcommand{\dd}{\mathrm{d}}

\newcommand{\Tr}{\mathrm{Tr}}

\newcommand{\rr}{\ensuremath{\mathbf{r}}}

\newcommand{\LL}{\ensuremath{\mathcal{L}}}
\newcommand{\VV}{\ensuremath{\mathcal{V}}}
\newcommand{\PP}{\ensuremath{\mathcal{P}}}
\newcommand{\QQ}{\ensuremath{\mathcal{Q}}}
\newcommand{\DD}{\ensuremath{\mathcal{D}}}
\newcommand{\HH}{\ensuremath{\mathcal{H}}}
\newcommand{\Yb}{$^{171}$Yb$^+~$}

\begin{document}

\title{Realising a quantum absorption refrigerator with an atom-cavity system}
\author{Mark T. Mitchison}
\email{marktmitchison@gmail.com}
\affiliation{Quantum Optics and Laser Science Group, Blackett Laboratory, Imperial College London, London SW7 2BW, United Kingdom}
\affiliation{Clarendon Laboratory, University of Oxford, Parks Road, Oxford OX1 3PU, United Kingdom}
\author{Marcus Huber}
\affiliation{Group of Applied Physics, University of Geneva, 1211 Geneva 4, Switzerland}
\affiliation{Departament de F\'{i}sica, Universitat Aut\`{o}noma de Barcelona, E-08193 Bellaterra, Spain}
\author{Javier Prior}
\affiliation{Universidad Polit\'{e}cnica de Cartagena, Paseo Alfonso XIII, 30203 Cartagena, Spain}
\author{Mischa P. Woods}
\affiliation{University College London, Department of Physics \& Astronomy, London WC1E 6BT, United Kingdom}
\affiliation{QuTech, Delft University of Technology, Lorentzweg 1, 2611 CJ Delft, Netherlands}
\author{Martin B. Plenio}
\affiliation{Institut f\"{u}r Theoretische Physik, Albert-Einstein Allee 11, Universit\"{a}t Ulm, 89069 Ulm, Germany}

\begin{abstract}
An autonomous quantum thermal machine comprising a trapped atom or ion placed inside an optical cavity is proposed and analysed. Such a machine can operate as a heat engine whose working medium is the quantised atomic motion, or as an absorption refrigerator which cools without any work input. Focusing on the refrigerator mode, we predict that it is possible with state-of-the-art technology to cool a trapped ion almost to its motional ground state using a thermal light source such as sunlight. We nonetheless find that a laser or similar reference system is necessary to stabilise the cavity frequencies. Furthermore, we establish a direct and heretofore unacknowledged connection between the abstract theory of quantum absorption refrigerators and practical sideband cooling techniques. We also highlight and clarify some assumptions underlying several recent theoretical studies on self-contained quantum engines and refrigerators. Our work indicates that cavity quantum electrodynamics is a promising and versatile experimental platform for the study of autonomous thermal machines in the quantum domain.
\end{abstract}
\maketitle

\section{Introduction}

Cooling of atomic motion is an essential precursor to a broad range of experiments with trapped atoms and ions. The development of laser cooling techniques represents a major achievement of late 20$^\mathrm{th}$ century physics, having enabled spectacular advances in our ability to study and manipulate quantum systems in the laboratory. More recently, researchers have begun to investigate quantum absorption refrigerators: machines which can cool using only a source of heat, without the need for work supplied by an external field, such as a laser. However, the application of such devices to practical tasks in quantum technology, such as the refrigeration of trapped atoms, has barely been studied thus far. 

Absorption refrigerators belong to the class of autonomous thermal machines (ATMs), i.e.\ those which operate without external control or work. Although such devices have existed since the dawn of thermodynamics, they have attracted renewed interest from quantum physicists \cite{Kosloff2014arpc,Goold2016jpa} for several reasons. ATMs are attractive to the theorist because they dispense of the need for externally supplied work, whose precise definition in a quantum setting is debatable \cite{Campisi2011rmp,Horodecki2013a,Skrzypczyk2014}. Furthermore, naturally occurring biological ATMs, such as photosynthetic complexes, have been found to exhibit quantum coherent dynamics \cite{Huelga2013cp,Killoran2015jcp}. Most importantly, ATMs potentially offer tremendous practical reductions in energy expenditure. This is because work is usually performed on a quantum system using a coherent field, which consumes a macroscopic quantity of power merely in order to control microscopic degrees of freedom. Conversely, an absorption chiller may be powered by ubiquitous sources of thermal energy, such as excess heat generated by another process, or indeed sunlight. In principle, absorption refrigerators can therefore function at no additional energy cost beyond that required to build the device in the first place.

In the recent theoretical literature, a number of quantum absorption refrigerator (QAR) models have been proposed and studied \cite{Linden2010prl, Levy2012prl, Levy2012pre, Gelbwaser2014pre, Correa2014pre, Silva2015pre, Wang2015pre, Leggio2015pra, Doyeux2016pre}. In many cases, these models exhibit new, occasionally controversial, behaviours that are absent from their classical counterparts. Quantum correlations have variously been argued to increase efficiency \cite{Brunner2013}, or to play no operational role \cite{Correa2013pre}, depending on the model considered. Performance advantages due to non-equilibrium \cite{Correa2014, Leggio2015pra} or spatially correlated \cite{Doyeux2016pre} states of the heat reservoirs have been discussed. Interesting effects have also been predicted in the transient regime, where quantum oscillations may facilitate fast cooling below the steady-state temperature \cite{Mitchison2015njp,Brask2015pre}. On the other hand, only a handful of concrete experimental proposals for QARs have been put forward \cite{Chen2012epl, Mari2012prl, Venturelli2013prl}. It is therefore of interest to explore other systems which could realise QARs, in order to better understand their physical limitations and capabilities in a practical context. 

With this goal in mind, we introduce and study a novel design for a QAR in a cavity optomechanical set-up using trapped atoms \cite{Kimble1999prl,McKeever2003prl,Birnbaum2005nat,Fortier2007prl} or ions \cite{Russo2009apb,Sterk2012pra,Stute2012apb}. We identify the necessary ingredients for the construction of such a machine, which include high-finesse optical cavities whose line width is smaller than the oscillation frequency of the trapped atom, although neither strong coupling nor high cooperativity are needed. Assuming that the necessary conditions are met, we show that driving the cavity with incoherent thermal light leads to significant cooling of the atomic motion. In particular, we predict that sunlight may be used to cool a trapped ion down to its motional ground state with high fidelity. Interestingly, we find that a laser is indispensable even in this context, however its role in our set-up is not to perform work, but rather to provide a stable frequency reference. Furthermore, the refrigeration cycle in our scheme can be easily understood by direct analogy with laser sideband cooling. This establishes a direct connection between the abstract theory of QARs and practical cooling techniques that are well known in atomic physics. 

Trapped-ion systems have already proved to be a successful experimental testing ground for quantum thermodynamics, with the recent proposal and subsequent realisation of a heat engine using a single ion as the working medium \cite{Abah2012prl, Rossnagel2015arXiv}. However, it is not yet clear how useful work could be extracted from such a device. In contrast, we focus on cooling, a task that has a clear practical application in quantum state preparation.  We also note that several practical laser cooling schemes for trapped atoms using optical cavities already exist \cite{Horak1997prl,Maunz2004nat,Nuszmann2005natphys,Leibrandt2009prl}. Our proposal differs from all of these, primarily because the energy for cooling is provided by a thermal source such as sunlight. Finally, we mention a closely related recent article \cite{Mari2012prl}, describing a QAR comprising a nano-mechanical oscillator interacting with a pair of optical resonators. Our scheme works by a similar mechanism, but relies on a different interaction Hamiltonian, and is capable of achieving temperatures lower by many orders of magnitude, even with much less intense thermal light input. We aim to elucidate the connections between schemes such as that of Ref.~\cite{Mari2012prl}, conventional sideband cooling, and other QAR models discussed in the recent theoretical literature.

This paper is arranged as follows. In Section~\ref{3bodyQAR} we outline the general theory of the three-body quantum absorption refrigerator and introduce some basic concepts. Section~\ref{singleCavity} is concerned with an idealised model involving a trapped atom inside a single optical cavity, which serves to illustrate some of the physical principles and limitations in a simplified context. In Section~\ref{doubleCavity} we describe our main proposal to build an absorption refrigerator using trapped atoms or ions inside a pair of optical cavities, and analyse the performance of the refrigerator in detail. We discuss our results and conclude in Section~\ref{conclusion}. Mathematical details are provided in the appendices.

\section{Three-body quantum absorption refrigerator}
\label{3bodyQAR}

\begin{figure}[t]
\includegraphics[width = 0.7\linewidth]{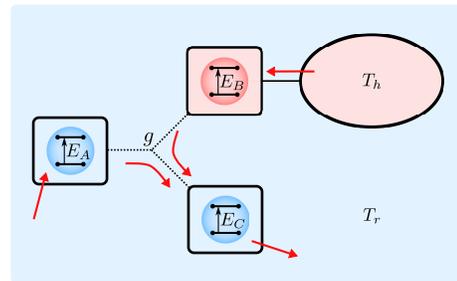}
\caption{Illustration of a quantum absorption refrigerator comprising three mutually interacting subsystems, each possessing a transition of energy $E_j$, such that $E_A + E_B = E_C$. Subsystem $B$ couples to a heat reservoir at temperature $T_h$ while the others interact with the environment at room temperature $T_r$. Red arrows show the  direction of steady-state heat flow through the machine. \label{fridgeCartoon}}
\end{figure}

In this section we introduce some fundamental concepts which form the basis for our work. In particular, we provide a concise, general exposition of the abstract model of the three-body QAR introduced in Refs.~\cite{Linden2010prl,Levy2012prl}. We also take this opportunity to introduce the useful concepts of virtual qubits and virtual temperatures, and to justify the figures of merit used to characterise refrigerator performance in subsequent sections. Readers familiar with the literature on QARs may wish to skip to Section~\ref{singleCavity}, where we specialise to atom-cavity systems. Unless otherwise indicated, we use units of energy and temperature such that $\hbar = 1$ and $k_B = 1$.

\subsection{Description of the model}

The three-body quantum absorption refrigerator comprises three subsystems with Hamiltonian 
\begin{equation}
\label{Hloc}
H = \sum_{j=A,B,C} H_j,
\end{equation}
where the operators $H_j$ act non-trivially on subsystem $j$ only. Subsystem $A$ is the body to be cooled. We assume that it has an equally spaced energy spectrum with level splitting $E_A$, i.e.\
\begin{equation}
\label{HaEquidistant}
H_A = E_A \sum_{n=0}^{D-1} n \ketbra{n}{n},
\end{equation}
where $D$ is the local Hilbert space dimension (possibly infinite). This general form may describe a qubit, a spin or a harmonic oscillator. We also assume that $H_B$ and $H_C$ each possess at least one pair of eigenstates differing in energy by $E_j$, such that
\begin{equation}
\label{resonanceCondition}
E_A + E_B = E_C.
\end{equation}
The form of $H_B$ and $H_C$ is otherwise arbitrary. The subsystems are coupled together by the three-body interaction
\begin{equation}
\label{general3bodyTerm}
V = g \left ( L_A L_B L_C^\d + L_A^\d L_B^\d L_C\right ),
\end{equation}
where $g$ is the interaction energy and $L_j$ is a lowering operator connecting pairs of Hamiltonian eigenstates separated by an energy $E_j$, i.e.\
\begin{equation}
\label{loweringOperatorJ}
[H, L_j] = - E_j L_j,
\end{equation}
while $L_j^\d$ is the corresponding raising operator. The condition \eqref{resonanceCondition} ensures that $[H,V] = 0$, so that the interaction \eqref{general3bodyTerm} enacts resonant transitions between degenerate energy eigenstates of $H$.

Cooling is achieved by coupling subsystem $B$ to a hot thermal bath at temperature $T_h > T_r$, while subsystems $A$ and $C$ remain coupled to the environment at temperature $T_r$. Energy exchange between the subsystems then allows heat to naturally flow from the hot reservoir to the colder environment. However, due to the specific form of the interaction \eqref{general3bodyTerm}, subsystem $C$ can only absorb a quantum of energy from the hot body $B$ by simultaneously absorbing energy from $A$, thus leading to cooling. The heat flow through the refrigerator is illustrated in Fig.~\ref{fridgeCartoon}.

\subsection{Virtual qubits and virtual temperatures}

The concepts of virtual qubits and virtual temperatures provide a convenient and intuitive way to analyse autonomous thermal machines \cite{Brunner2012pre}. A virtual qubit is a pair of states in the composite Hilbert space of subsystems $B$ and $C$ which directly couples to the target subsystem $A$. By choosing the parameters of the system appropriately, the virtual qubit can be placed at an effective virtual temperature which may be lower than $T_r$. The operation of the refrigerator can then be understood as a simple thermalisation process between $A$ and the virtual qubit. 

To make this notion explicit, we observe that the interaction Hamiltonian \eqref{general3bodyTerm} can be written as
\begin{equation}
\label{ThreebodyInteractionVirtual}
V = g \left ( L_A L_v^\d + L_A^\d L_v \right ),
\end{equation}
where $L_v  = L_B^\d L_C$. Eqs.~\eqref{resonanceCondition} and \eqref{loweringOperatorJ} together imply that 
\begin{equation}
\label{virtualLoweringOperator}
[H, L_v] = -E_A L_v.
\end{equation}
This means that $L_v$ is a lowering operator connecting pairs of states differing by an energy $E_A$ in the composite Hilbert space of $B$ and $C$. Each of these pairs of states is called a virtual qubit. The interaction~\eqref{ThreebodyInteractionVirtual} then describes resonant energy exchange between the virtual qubits and $A$.

When each subsystem is at thermal equilibrium with its respective bath, the populations of the virtual qubit states are thermally distributed at a virtual temperature
\begin{equation}
\label{virtualTemperature}
T_v = \frac{ E_A}{E_C/T_r -  E_B/T_h}.
\end{equation}
That is to say, each pair of virtual qubit states is populated in the ratio $\ee^{-E_A/ T_v}$. As long as the parameters of the refrigerator are chosen so that $T_v < T_r$, subsystem $A$ will be pushed towards a lower temperature as it equilibrates with the virtual qubits under the interaction~\eqref{ThreebodyInteractionVirtual}. This effect is counteracted by the thermalising influence of the reservoir interacting with $A$, thus establishing a heat current flowing from the environment surrounding $A$ into the refrigerator. 
See Ref.~\cite{Brunner2012pre} for a more complete discussion of virtual qubits and temperatures. 
 
To conclude this section, we briefly mention that if the assumption that $T_h > T_r$ is relaxed, one can arrange for $T_v$ to take any value by adjusting the bath temperatures and energy splittings. If $T_v > T_r$, then the steady-state temperature of $A$ is increased and the system operates as a heat pump. On the other hand, if $T_v < 0$ the machine tries to induce population inversion in the state of $A$, which can be thought of as the quantum analogue of a classical heat engine lifting a weight \cite{Brunner2012pre}. In the following, we restrict our attention to the absorption chiller mode, where $T_h \gg T_r$. However, our results could be applied equally well to the construction and study of autonomous quantum heat pumps and engines.

\subsection{Figures of merit}

In order to analyse the performance of a refrigerator, one must choose figures of merit. The appropriate figure of merit depends on the problem at hand, as we now explain. 

\subsubsection{Coefficient of performance and cooling power}

From one viewpoint, the refrigerator can be seen as a device which extracts heat from the environment surrounding subsystem $A$. Thus, the refrigerator performance is characterised by the stationary heat currents flowing to and from the reservoirs. The cooling power $\dot{Q}_A$ gives the heat current into the refrigerator from the environment of $A$, while $\dot{Q}_B$ gives the input power corresponding to the heat current flowing in from the hot reservoir. Therefore, the relevant figure of merit is the coefficient of performance $\epsilon = \dot{Q}_A/\dot{Q}_B$. 

Note that this point of view makes sense only if the reservoirs connected to subsystems $A$ and $C$ are considered as separate entities. If subsystems $A$ and $C$ are in fact connected to the same environment, the net effect of the machine is simply to dump $\dot{Q}_B$ energy per unit time from the hot reservoir into this environment. 

Assuming that the reservoirs connected to $A$ and $C$ are independent, we can estimate the coefficient of performance using the equations of motion for the mean local energies of subsystems $A$ and $B$, viz.
\begin{equation}
\label{meanEnergyEqnA}
\dt{\langle H_j\rangle} = \dot{Q}_j + \ii g E_j \avg{ L_A L_B L_C^\d - L_A^\d L_B^\d L_C },
\end{equation}
for $j = A,B$, where the second term on the right-hand side (RHS) follows from the Heisenberg equation generated by the interaction Hamiltonian \eqref{general3bodyTerm}. In the stationary state, the derivatives of the mean energies vanish, leading to the following simple expression for the coefficient of performance:
\begin{equation}
\label{COP}
\epsilon = \frac{E_A}{E_B}.
\end{equation}
We see that the coefficient of performance grows without limit as $E_A$ is increased while holding $E_B$ fixed (assuming that Eq.~\eqref{resonanceCondition} is always satisfied). It is important to note that our approximate analysis ignores the contribution of the interaction energy to the heat currents, and is therefore only strictly correct in the weak-coupling limit of vanishingly small $g$ \cite{Correa2013pre}.

\subsubsection{Achievable temperature and cooling time}

In the above scenario, one uses a microscopic machine to cool a \textit{macroscopic} body, namely the reservoir connected to $A$. Perhaps a more appropriate application of a quantum refrigerator is to cool a \textit{microscopic} system, namely subsystem $A$ itself. From this viewpoint, the most important figure of merit is the achievable temperature (or more generally, the achievable energy and entropy) of subsystem $A$ \cite{Mitchison2015njp}.

The achievable steady-state temperature can be estimated from the virtual temperature given by Eq.~\eqref{virtualTemperature}. This takes its minimal value when the temperature of the hot bath is large, from which we find that
\begin{equation}
\label{virtualTempThInf}
\lim_{T_h\to\infty} T_v = \frac{E_A}{E_C} T_r  = \frac{\epsilon}{1+\epsilon} T_r,
\end{equation}
where we have used Eqs.~\eqref{resonanceCondition} and \eqref{COP} to rewrite the virtual temperature in terms of the coefficient of performance $\epsilon$. We find that the virtual temperature is minimised when $\epsilon$ is small. This illustrates that the standard thermodynamic measures of steady-state refrigerator performance are essentially irrelevant when the task at hand is to cool a quantum system having a finite energy.

Since the refrigerator is out of equilibrium, the thermodynamic temperature of $A$ may not be strictly defined. For our purposes, it is sufficient to adopt the mean energy $\avg{H_A}$ as a figure of merit, rather than the temperature. This also provides an adequate measure of entropy, since the von Neumann entropy of a state with mean energy $\avg{H_A}$ is upper-bounded by that of a Gibbs state having the same mean energy. If the cooling is subject to time constraints, the relaxation time (the time taken for $\avg{H_A}$ to reach its stationary value) is also a measure of performance. However, the relaxation time is a non-universal figure of merit since it may depend on the initial conditions.

Throughout the remainder of this article we adopt the present framework, where the objective is to cool subsystem $A$. This viewpoint is particularly appropriate for quantum technology applications. Here, the motivation for cooling a quantum system is typically to maximise the efficiency of subsequent control operations by reducing uncertainty over the initial conditions, i.e. by minimising the entropy of the quantum system. We therefore neglect traditional efficiency measures such as the coefficient of performance $\epsilon$, choosing rather to focus on the mean energy of the subsystems constituting the refrigerator, in particular that of the target body $A$. 

\section{Single-cavity configuration}
\label{singleCavity}

In this section we introduce an idealised model of a quantum absorption refrigerator comprising a trapped atom inside a single optical cavity. In order to simplify the analysis, several details are disregarded in this section. Nevertheless, this simplified model is sufficient to illustrate the physical principles involved. The present toy model also has the advantage of making the connection with laser sideband cooling obvious, while demonstrating some of the practical limitations which arise in the cavity quantum electrodynamics (CQED) setting. 

\subsection{Description of the system}
\label{singleCavityDescription}

\begin{figure}
\includegraphics[width=0.7\linewidth]{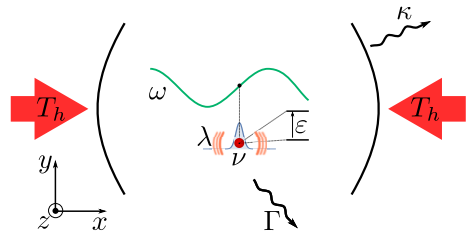}
\caption{Schematic of the single cavity set-up. The trap potential minimum coincides with an electric-field node of a cavity mode with frequency $\omega = \varepsilon-\nu$. Dissipation leads to line widths $\Gamma$ and $\kappa$ for the electronic transition and the cavity mode, respectively, while $\lambda$ is the intrinsic heating rate of the atomic motion in the trap.\label{singleCavityCartoon}}
\end{figure}

We consider a single atom or ion of mass $M$ confined in the $x$ direction by a harmonic potential with oscillation frequency $\nu/2\pi$. The atom is assumed to possess a pair of relevant internal electronic states $\ket{\downarrow}$ and $\ket{\uparrow}$ separated by an energy $\varepsilon$. The trap is placed inside an optical cavity whose axis is aligned in the $x$ direction. The minimum of the harmonic potential is placed at a node of the electric field of a cavity mode with frequency $\omega/2\pi$ chosen such that $\omega = \varepsilon - \nu$. We assume that $\omega\sim \varepsilon$ and $\varepsilon,\omega \gg \nu$, as appropriate for typical optical and vibrational frequencies. The geometry of the problem is depicted schematically in Fig.~\ref{singleCavityCartoon}.

The free Hamiltonian of the system is 
\begin{equation}
\label{H0}
H_1 =  \nu a^\d a + \omega b^\d b + \varepsilon \sigma^+ \sigma^- ,
\end{equation}
where the bosonic ladder operators $a^\d$ and $b^\d$ respectively create motional quanta (phonons) and light quanta (photons), while $\sigma^- = \ketbra{\downarrow}{\uparrow} = (\sigma^+)^\d$ is the atomic lowering operator. We have assumed that all other electronic states and vibrational or cavity modes are far off-resonant and can be neglected. 

The interaction between the atom and the cavity field in the dipole approximation reads as
\begin{equation}
\label{singleCavityInteraction}
V_1 =  g \sin \left [ \eta \left (a + a^\d \right ) \right ]  \left ( b + b^\d \right )\left ( \sigma^- + \sigma^+ \right ),
\end{equation}
where $g$ is the cavity coupling constant and the Lamb-Dicke parameter is defined as $\eta = \omega/\sqrt{2M c_0^2 \nu}$, with $c_0$ the speed of light in vacuum. The form of the interaction~\eqref{singleCavityInteraction} reflects the symmetry of the problem when the harmonic potential minimum coincides exactly with an electric field node. The electric field operator then changes sign under a parity transformation of the atomic centre-of-mass coordinate. This gives rise to a selection rule allowing only transitions between motional states of opposite parity. Since the vibrational energy eigenstates have definite parity, the absorption or emission of a photon must therefore be accompanied by a change in the number of phonons. 

We now show that the system approximately realises a QAR, and estimate its virtual temperature. We work in the limit $\eta \ll 1$, which requires that the cavity mode wavelength is much larger than the characteristic length scale of atomic motion. We can therefore invoke the Lamb-Dicke approximation (LDA) and expand Eq.~\eqref{singleCavityInteraction} to first order in $\eta$. We also make the rotating wave approximation (RWA) by discarding counter-rotating terms at optical frequency, leading to
\begin{equation}
\label{singleCavityInteractionRWA}
V_1 \approx g\eta \left ( a + a^\d \right ) \left ( b\sigma^+ + b^\d \sigma^- \right ).
\end{equation}
Assuming that terms counter-rotating at frequency $\pm 2\nu$ can also be neglected, we finally obtain
\begin{equation}
\label{singleCavityInteractionSecondRWA}
V_1 \approx  g \eta\left ( a  b \sigma^+ + a^\d  b^\d\sigma^- \right ),
\end{equation}
We therefore find under these assumptions that the system exhibits a three-body interaction of the type \eqref{general3bodyTerm}. Note that the approximation leading to Eq.~\eqref{singleCavApproxTv} is valid only when $g\eta\ll 2\nu$. Eq.~\eqref{singleCavApproxTv} also assumes that the line widths of relevant transitions are much smaller than the trap frequency $\nu$, as discussed in detail in Section \ref{singlePerformance}.

The heat reservoir is provided by coupling thermal light at a high temperature $T_h$ into the cavity resonator. Assuming that the thermal light source is well collimated, the electronic transition couples only to the ambient radiation field at room temperature $T_r \ll T_h$, which leads to spontaneous emission. Meanwhile, the motion of the atom undergoes intrinsic heating in the trap, for example due to fluctuations of the trapping potential.

The virtual qubit states in the machine are the pairs $\{\ket{n_b,\downarrow},\ket{n_b-1,\uparrow}\}$, where $\ket{n_b}$ denotes a Fock state with $n_b$ quanta in the cavity mode. The virtual temperature of the refrigerator is therefore
\begin{equation}
\label{singleCavityVirtualTemp}
T_v = \frac{\nu}{\varepsilon/T_r - \omega/T_h}.
\end{equation}
Since $T_h \gg T_r$ and $\nu\ll \varepsilon$, we find that
\begin{equation}
\label{singleCavApproxTv}
\frac{T_v}{T_r} \approx \frac{\nu}{\varepsilon}.
\end{equation}
The ratio of frequencies is typically $\nu/\varepsilon \sim 10^{-8}$ or less, implying that very low virtual temperatures are achievable.

\begin{figure}
\includegraphics[width=0.7\linewidth]{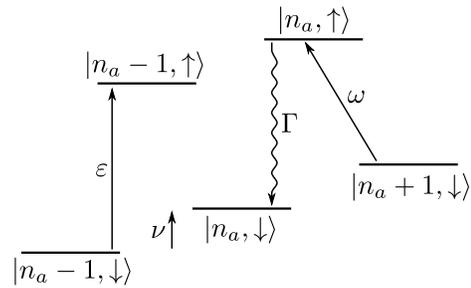}
\caption{Level scheme showing the manifold of electronic and vibrational states for the single-cavity configuration. Thermal cavity photons resonant with the red sideband transition are absorbed, then the electronic state is reset by spontaneous emission, driving the system down the ladder of vibrational states. \label{singleCavityLevels}}
\end{figure}

The refrigerator operation can be understood by a straightforward analogy with laser sideband cooling. The optical cavity behaves as a filter which singles out frequencies close to the red sideband $\omega_\mathrm{red} = \varepsilon - \nu$. Pumping the cavity with thermal light increases the number of photons with the correct frequency to drive the red sideband transition. Spontaneous emission then resets the electronic state, completing the cooling cycle (see Fig.~\ref{singleCavityLevels}). So long as the blue sideband frequency $\omega_\mathrm{blue} = \varepsilon + \nu$ is far off-resonant, the absorption of thermal cavity photons drives the motion towards its ground state. 

\subsection{Master equation}

In order to study the dynamics of the model, we employ a quantum master equation for the density operator $\rho$ of the form
\begin{equation}
\label{masterEquationSingle}
\dt{\rho} =  -\ii[H_1+V_1,\rho] + \sum_{j=a,b,\sigma}\LL_j\rho.
\end{equation}
The superoperators $\LL_j$ are dissipative contributions due to the coupling of each subsystem to its respective reservoir. The term $\LL_a$ describes motional heating, $\LL_{b}$ corresponds to the thermal pumping of the cavity, while $\LL_\sigma$ relates to spontaneous emission.

Introducing the general notation for a Lindblad dissipator
\begin{equation}
\label{dissipator}
\mathcal{D}[L]\rho = L\rho L^\d - \frac{1}{2}\{L^\d L,\rho\},
\end{equation}
the motional heating is described by 
\begin{equation}
\label{La}
\LL_a = \lambda(1 + \bar{n}_a^{-1}) \DD[a] + \lambda \DD[a^\d],
\end{equation}
where $\bar{n}_a = (\ee^{\nu/T_r} - 1)^{-1}$ is the equilibrium phonon number at room temperature. Since $\bar{n}_a^{-1} \approx 0$ to an excellent approximation under typical laboratory conditions, the dissipator \eqref{La} describes approximately linear growth of the phonon number at a constant heating rate $\lambda$. Pumping of the cavity with thermal light is described by
\begin{equation}
\label{Lb}
\LL_b = 2\kappa(1+\bar{n}_b)\DD[b] + 2\kappa\bar{n}_b\DD[b^\d],
\end{equation}
where $\kappa$ is the cavity line width and $\bar{n}_b = ( \ee^{\omega/T_h} - 1)^{-1}$  is the equilibrium photon number at temperature $T_h$. For simplicity, we have assumed here that the thermal driving is applied to both sides of the cavity. Spontaneous emission is described by the Liouvillian
\begin{align}
\label{Lsig}
\LL_\sigma  = &\; \Gamma (1 + \bar{n}_\sigma) \int_{-1}^{1}\dd u\, \Pi(u) \DD[\ee^{\ii\eta u (a+a^\d)} \sigma^-] \notag \\
& +\,\Gamma  \bar{n}_\sigma \int_{-1}^{1}\dd u\, \Pi(u) \DD[\ee^{-\ii\eta u (a+a^\d)} \sigma^+] ,
\end{align}
with $\bar{n}_\sigma = (\ee^{\varepsilon/T_r} - 1)^{-1}$, while $\Pi(u)$ is the angular distribution of emitted photons as a function of $u = \cos\theta$, where $\theta$ is the angle subtended from the $x$ axis by the photon wave vector. Note that $\bar{n}_\sigma \approx 0$ at optical frequencies and room temperature, and therefore the absorption term on the second line of Eq.~\eqref{Lsig} is typically negligible.

\subsection{Line broadening and other constraints}
\label{singlePerformance}

Unfortunately, the single-cavity refrigerator suffers from several severe practical limitations. The most important of these is due to line broadening. The picture illustrated in Fig.~\ref{singleCavityLevels}, described by the Hamiltonian~\eqref{singleCavityInteractionSecondRWA}, is valid when the sideband transitions are ``sharp", in the sense of having a well-defined frequency. However, thermal dissipation implies some unavoidable energy uncertainty due to the finite lifetime of the states involved. Significant cooling is only possible in the sideband-resolved regime, where the frequency uncertainty of the relevant transitions is much less than $\nu$. Otherwise, line broadening brings the blue sideband transition partially onto resonance, leading to heating rather than cooling. In particular, this means that we must have $\lambda,\kappa, \Gamma < \nu$ for effective refrigeration.

\begin{figure}
\includegraphics[width = 0.8\linewidth]{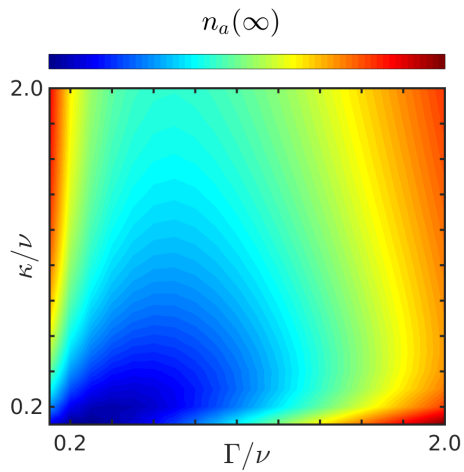}
\caption{Qualitative dependence of the steady-state phonon occupation $n_a(\infty)$ on the spontaneous emission rate $\Gamma$ and the cavity decay constant $\kappa$, with $\eta = 0.05$, $g = \nu$, $\bar{n}_b = 10^{-3}$ and $\lambda = 0$. The scale for $n_a(\infty)$ goes from blue (cold) to red (hot).\label{singleColourPlot}}
\end{figure}

In order to illustrate the effect of line broadening, we compute the steady-state phonon occupation $n_a(\infty) = \Tr[a^\d a \rho_\infty]$, where $\rho_\infty$ is the stationary quantum state satisfying $\dd\rho_\infty/\dd t = \LL\rho_\infty = 0$, with $\LL$ given by the RHS of Eq.~\eqref{masterEquationSingle}. The problem is simplified by taking the interaction Hamiltonian~\eqref{singleCavityInteractionRWA} under the LDA and RWA, and making the approximation $\bar{n}_a^{-1} \approx 0 \approx \bar{n}_\sigma$, leaving just five free parameters governing the phonon population dynamics:  $\lambda$, $\eta$, $\kappa$, $\bar{n}_b$ and $\Gamma$. We compute the stationary state by representing $\LL$ as a matrix and solving the eigenvalue equation $\LL\rho_\infty = 0$. The integral in Eq.~\eqref{Lsig} is numerically approximated by a trapezoidal rule. We take $\Pi(u) = 3(1+u^2)/8$, as appropriate for a point dipole aligned perpendicularly to the cavity axis. Sampling a grid of 100 evenly spaced points in the interval $u \in [-1,1]$ is sufficient to obtain convergence. The resulting Liouvillian matrix has low sparsity and is therefore challenging to diagonalise, which limits the achievable Hilbert space dimension considerably. We use 21 phonon states and 4 cavity photon states in total. The results are therefore quantitatively inaccurate, but suffice to obtain qualitative trends.

The qualitative dependence of $n_a(\infty)$ on the dissipation rates $\kappa$ and $\Gamma$ is plotted for some example parameters in Fig.~\ref{singleColourPlot}. We see that the optimum operating regime is $\kappa,\Gamma\ll \nu$, as expected. The performance deteriorates rapidly as $\Gamma$ or $\kappa$ is increased above the trap frequency $\nu$. Increasing the spontaneous emission rate has a particularly adverse effect, because the recoil momentum of emitted photons leads to further motional heating (see Eq.~\eqref{Lsig}). This is highly problematic, because the spontaneous emission rate in atomic two-level systems is fixed by Nature, and may be much larger than a typical vibrational frequency, on the order of tens or hundreds of megaherz.

The deleterious effect of line broadening is worsened when the trapping potential minimum is not placed exactly on the electric-field node of the cavity. Outside of the sideband-resolved regime, we have found that the system is remarkably sensitive to small misalignments of the trapping potential: displacements of a few nanometres away from the cavity field node lead to an almost complete disappearance of the cooling effect. This can be understood as follows. Away from the node, cavity photons may be emitted and absorbed without affecting the vibrational state of the atom. Photon absorption in particular depletes the cavity field and reduces the effective temperature of the hot reservoir. Although these transitions, which occur at the so-called carrier frequency $\varepsilon$, are off-resonant in principle, they become important when levels are broadened.

Finally, in order to enforce the resonance condition $\omega = \varepsilon - \nu$, it is necessary to stabilise the length of the cavity to prevent frequency drift. This may be achieved, for example, by continuously driving the cavity with a laser field and using the Pound-Drever-Hall technique \cite{Black2001ajp}. Importantly, this stabilisation can be performed using other polarisation modes or different cavity harmonics from those directly relevant for the refrigerator's operation. Of course, this use of an external laser field means that the machine is not truly autonomous. However, the laser does not supply any work used directly for cooling. Rather, its role is to provide a stable frequency reference. 

\section{Crossed-cavity configuration}
\label{doubleCavity}

In this section we describe a detailed model of a quantum absorption refrigerator comprising a trapped atom within a pair of perpendicular optical cavities. The purpose of the additional cavity is to ameliorate the adverse effects of spontaneous emission. We predict that such a machine powered by sunlight can cool a trapped ion to near its motional ground state, and explicitly delineate the parameter regime in which this is possible.

\subsection{Description of the model}

As in Section~\ref{singleCavityDescription}, we consider a harmonically trapped atom or ion of mass $M$, possessing a pair of electronic states $\ket{\downarrow}$ and $\ket{\uparrow}$ separated by energy $\varepsilon$. In this section we explicitly model the atomic motion in both the $x$ and $y$ directions, although it will shortly be shown that the $y$ coordinate decouples from the dynamics for our chosen configuration. For simplicity of presentation, we make the inessential assumption of equal oscillation frequencies in both the $x$ and $y$ directions, given by $\nu/2\pi$. 

\begin{figure}
\includegraphics[width=0.7\linewidth]{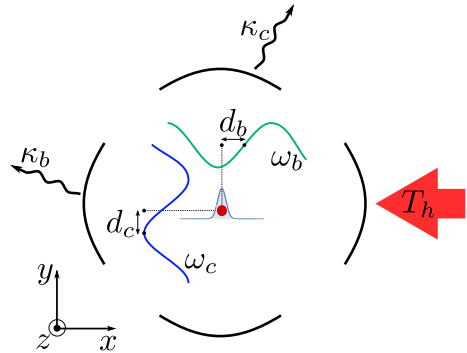}
\caption{Schematic of the crossed-cavity set-up. An atom is trapped close to an electric-field node of cavity mode $b$ and an anti-node of cavity mode $c$. Pumping mode $b$ with thermal light results in cooling of the atomic motion. A similar configuration was studied in Ref.~\cite{Plenio2002prl}, in the context of dissipative entanglement generation. \label{doubleCavityFig}}
\end{figure}

The atom is placed inside a pair of optical cavities $b$ and $c$, with axes aligned in the $x$ and $y$ direction, respectively. These cavities have relevant modes at frequencies $\omega_b/2\pi$ and $\omega_c/2\pi$. The minimum of the trap potential is placed a distance $d_b$ from a \textit{node} of the electric field in cavity $b$, and a distance $d_c$ from an \textit{anti-node} of cavity $c$. The geometry of the problem is indicated in Fig.~\ref{doubleCavityFig}. 

The free Hamiltonian of the system is
\begin{equation}
\label{doubleH0}
H_2 =  \nu a_x^\d a_x +\nu a_y^\d a_y + \omega_b b^\d b +  \omega_c c^\d c  + \varepsilon\sigma^+ \sigma^-,
\end{equation}
where $a^\d_x$ ($a_y$) creates motional excitations in the $x$ direction ($y$ direction), $b^\d$ ($c^\d$) creates photons in the cavity parallel to the $x$ axis ($y$ axis), and $\sigma^- = \ketbra{\downarrow}{\uparrow} = (\sigma^+)^\d$. The light-matter interaction Hamiltonian reads as
\begin{align}
\label{interactionHdouble}
V_2 = & \: g_b \sin \left[ \delta_b + \eta_b \left( a_x + a_x^\d \right) \right]\left (b + b^\d \right ) \left (\sigma^- + \sigma^+ \right ) \notag\\ & + g_c \cos \left [ \delta_c + \eta_c \left ( a_y + a_y^\d \right ) \right ]\left (c + c^\d \right ) \left (\sigma^- + \sigma^+ \right ),
\end{align}
where $g_j$ is the coupling constant, $\eta_j = \omega_j/\sqrt{2M c_0^2 \nu}$ is the Lamb-Dicke parameter, and $\delta_j = d_j\omega_j/c_0$ is the dimensionless misalignment for cavity $j = b,c$. As before, all other cavity and vibrational modes and electronic states are assumed be far off-resonant.

 Assuming that $\eta_j,\delta_j\ll 1$, we expand Eq.~\eqref{interactionHdouble} to first order in small quantities and make the RWA, which yields
\begin{align}
\label{doubleInteractionLDRWA}
V_2 \approx & \: \tilde{g}_b \eta_b\left( a_x + a_x^\d \right) \left (b\sigma^+ + b^\d\sigma^- \right ) + \tilde{g}_c \left (c\sigma^+ + c^\d\sigma^- \right )\notag \\ & + h_b\left (b\sigma^+ + b^\d\sigma^- \right ),
\end{align}
where $\tilde{g}_{b/c} = g_{b/c}\cos\delta_{b/c}$ and $h_b = g_b\sin \delta_b$. We see that to lowest order, the excitation of phonons in the $y$ direction is suppressed close to the anti-node of cavity $c$. The motion in the $y$ direction is therefore neglected from here on. In order to simplify the notation we also set  $a_x = a$ and $\eta_b = \eta$.

We demand that the cavities be tuned to two-photon resonance with the red sideband, $\omega_c - \omega_b = \nu$, yet detuned from the carrier by an amount $\Delta = \omega_c - \varepsilon$, where $\varepsilon\gg \lvert\Delta\rvert \gg g_{b/c}$. Direct excitation of the internal state of the atom, and the associated spontaneous emission, is thus strongly suppressed. However, due to the resonance condition $\omega_c = \omega_b + \nu$, the cavities can coherently exchange photons, assisted by the creation or destruction of phonons. In the following subsection, we show that this process is described by the effective interaction
\begin{equation}
\label{doubleEffInteraction}
V_\mathrm{eff} = k \left( abc^\d + a^\d b^\d c \right ),
\end{equation}
where $k = \tilde{g}_b\tilde{g}_c\eta/\Delta$. This obviously corresponds to the general form \eqref{general3bodyTerm}.

The refrigerator is powered by pumping cavity mode $b$ with hot thermal light at temperature $T_h$, while cavity mode $c$ couples to the radiation field at room temperature $T_r$. The virtual qubit states for this system are the pairs $\{\ket{n_b,n_c},\ket{n_b-1,n_c+1}\}$. The virtual temperature is given by
\begin{equation}
\label{virtualTempDouble}
T_v = \frac{\nu}{\omega_c/T_r - \omega_b/T_h} \approx \frac{\nu}{\omega_c} T_r,
\end{equation}
since $T_h\gg T_r$ and $\omega_c\gg \nu$, and we see again that very low virtual temperatures can be obtained. 

\begin{figure}
\includegraphics[width=0.7\linewidth]{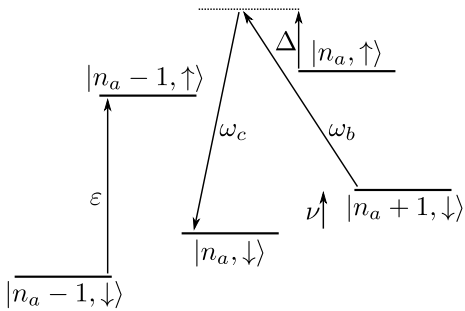}
\caption{Level scheme for the crossed-cavity configuration showing the direct analogy with Raman sideband cooling. Cooling on the red sideband occurs via a two-photon transition in which photons are exchanged between the two cavities. Spontaneous emission from the excited electronic state is suppressed by the detuning $\Delta$.\label{doubleCavityLevels}}
\end{figure}

The operation of the refrigerator can be understood by analogy with Raman laser sideband cooling (see Fig.~\ref{doubleCavityLevels}). Addressing the red sideband with a two-photon transition avoids populating the fast-decaying excited electronic state. The line width of the transition is therefore determined by the cavity decay rates, which in principle may be made much smaller than the spontaneous emission rate. This makes achieving the sideband-resolved regime a feasible prospect in this system. 

The cavity lengths must be actively stabilised in order to avoid frequency drift away from the resonance condition $\omega_c - \omega_b = \nu$. As described in Section~\ref{singlePerformance}, this stabilisation can be performed non-invasively using a laser. This makes the double-cavity refrigerator not truly autonomous. However, we emphasise again that the role of the laser is to provide a frequency reference, rather than to directly perform work used for cooling. 

\subsection{Master equation}
\label{doubleMEsec}

We write the density operator of the full system including the electronic degrees of freedom as $\chi$. This satisfies the master equation 
\begin{equation}
 \label{masterEquation}
 \dt{\chi} = -\ii[H_2 + V_2,\chi] + \sum_{j = a,b,c,\sigma} \mathcal{L}_j \chi,
 \end{equation} 
where $\LL_a$ and $\LL_\sigma$ are respectively defined by Eqs.~\eqref{La} and \eqref{Lsig}. The coupling of cavity $b$ to the external electromagnetic field is described by
\begin{equation}
\label{Lb}
\mathcal{L}_b = \kappa_b \left (2 + \bar{n}_b\right )\mathcal{D}[b] + \kappa_b\bar{n}_b \mathcal{D}[b^\d], 
\end{equation}
where $\kappa_{b}$ is the cavity line width and $\bar{n}_b = \left (\ee^{\omega_b/T_h} - 1 \right )^{-1}$ is the equilibrium photon number at temperature $T_h$. Eq.~\eqref{Lb} represents thermal driving applied to only one side of cavity $b$, while the other side couples to the vacuum (which approximates the electric field at room temperature). Cavity $c$ couples to the environment via the Liouvillian
\begin{equation}
\label{Lc}
\LL_c = 2\kappa_c (1 + \bar{n}_c) \mathcal{D}[c] + 2\kappa_c \bar{n}_c\DD[c^\d],
\end{equation}
where $\kappa_c$ is the corresponding line width, and $\bar{n}_c = (\ee^{\omega_c/T_r} - 1)^{-1}$, with $\bar{n}_c\approx 0$ for optical frequencies at room temperature.

We note that direct excitation of the electronic degrees of freedom is suppressed by the large detuning $\lvert\Delta\rvert \gg \tilde{g}_{b/c},h_b$. Furthermore, $\Gamma$ will typically be the largest dissipative frequency scale in the system, so that correlations between the electronic degrees of freedom and the rest of the system decay rapidly on the time scales relevant for the dynamics of the atomic motion. These assumptions enable us to simplify the model by adiabatically eliminating the excited electronic state within a Born-Markov approximation.

Using standard projection operator techniques \cite{QuantumNoise, Breuer2007book, VanKampen, Rivas2010njp}, a master equation describing the reduced density matrix $\rho(t) = \Tr_\sigma[\chi(t)]$ of the motional and cavity modes in the electronic ground state manifold is derived in Appendix \ref{appendixAdiabaticElimination}. The result is
\begin{equation}
\label{effMEdoubleMain}
\dt{\rho} = -\ii[H_{abc} + \delta H_{abc} + V_\mathrm{eff},\rho]+ \LL_\mathrm{se}\rho + \sum_{j=a,b,c}\LL_j \rho ,
\end{equation}
where
\begin{equation}
\label{HeffME}
H_{abc} =  \nu a^\d a + \omega_b b^\d b +  \omega_cc^\d c.
\end{equation}
The Hamiltonian $\delta H_{abc}$ is a small Lamb-shift contribution which renormalises the energy levels of $H_{abc}$. The interaction term is of the form
\begin{equation}
\label{VeffMain}
V_\mathrm{eff} = k\left (a b c^\d + a^\d b^\d c \right ).
\end{equation}
In the limit $\lvert\Delta\rvert \gg\Gamma$, the effective coupling constant is found to be $k = \tilde{g}_b\tilde{g}_c\eta/\Delta$.  The generator $\LL_\mathrm{se}$ describes additional dissipative processes due to spontaneous emission from the excited state, which are suppressed  by a factor of order $\Gamma/\Delta$ relative to the coherent coupling $k$. Full expressions for all parameters entering Eq.~\eqref{effMEdoubleMain} can be found in the appendix.

\begin{figure*}

     \begin{minipage}{0.4\linewidth}
          \flushleft \vspace{2mm}(a)  \vspace{-2mm} \begin{figure}[H]
			\includegraphics[scale = 0.25]{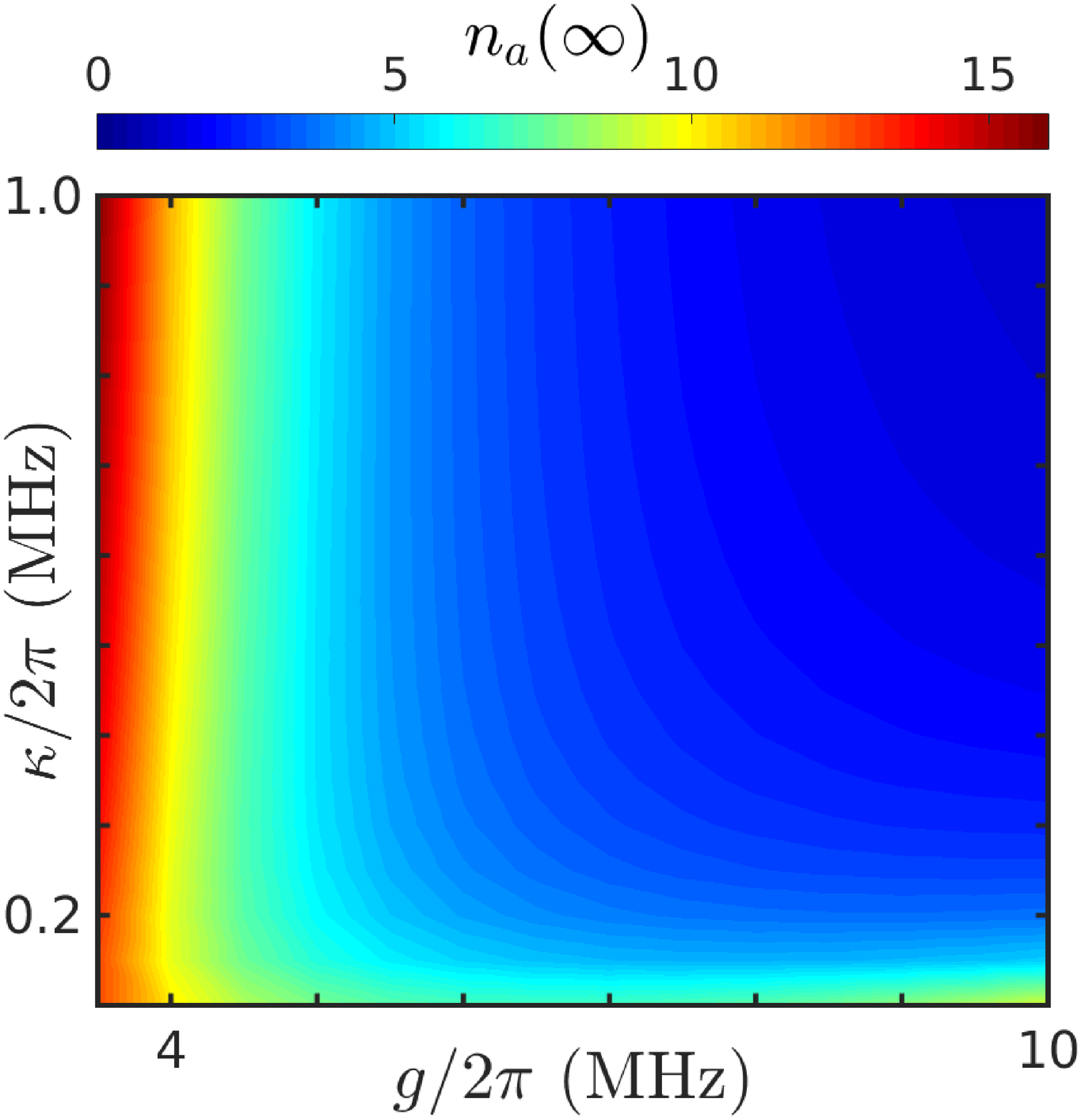}
          \end{figure}
      \end{minipage}
      \hspace{0.05\linewidth}
      \begin{minipage}{0.4\linewidth}
         \flushleft (b) \vspace{-2mm}\begin{figure}[H] 
              \includegraphics[scale = 0.28]{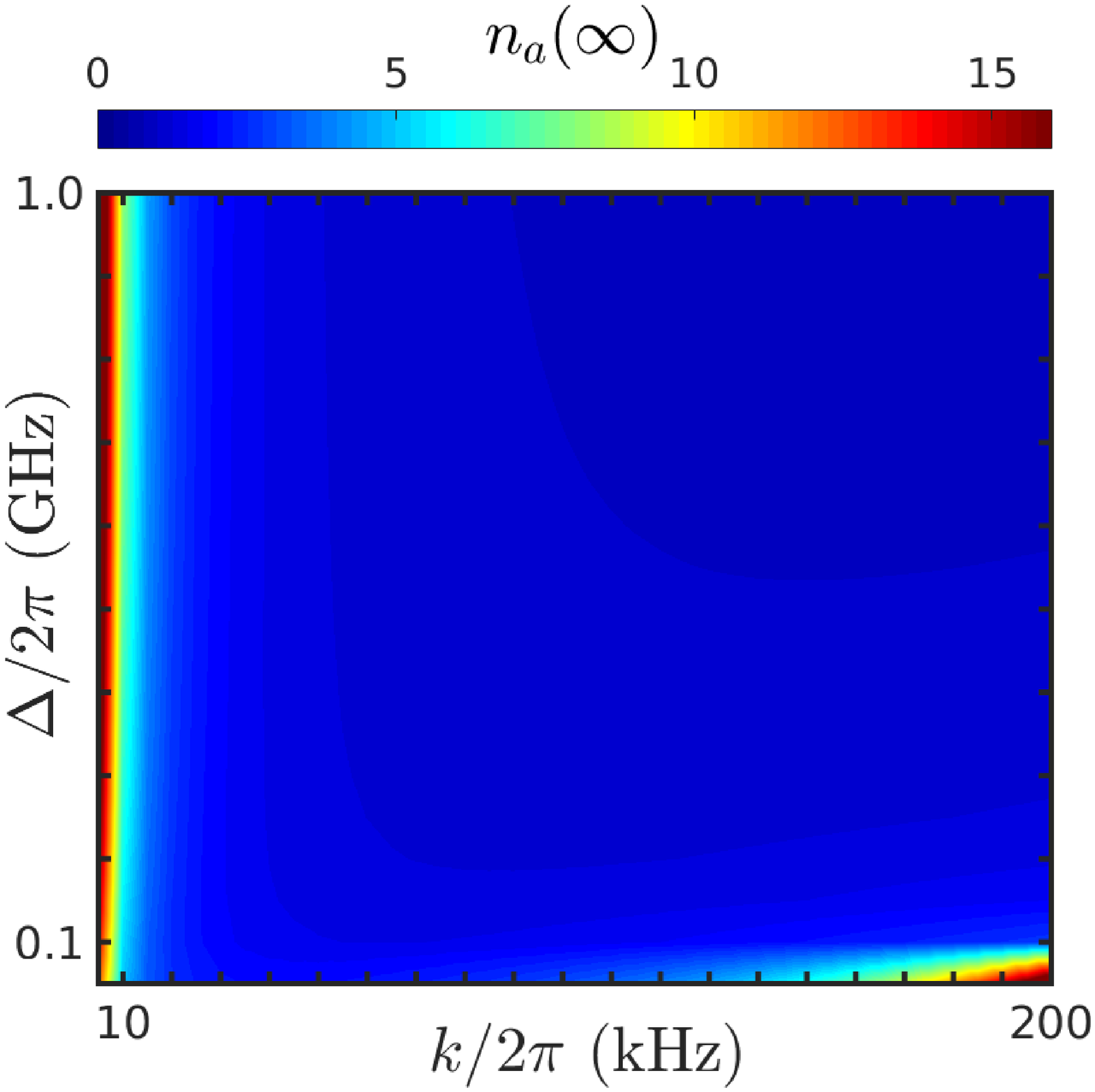}
          \end{figure}
          
      \end{minipage}

\caption{Steady-state phonon occupation of the crossed-cavity absorption refrigerator driven by sunlight. Parameters are given in Table~\ref{paramTab}, with (a) fixed $\Delta/2\pi = 100~$MHz and variable $g$ and $\kappa$, and (b) fixed $\kappa/2\pi = 0.5~$MHz and variable $k$ and $\Delta$. \label{doubleColourPlots}}
\end{figure*}

We now summarise the approximations underlying Eq.~\eqref{effMEdoubleMain}. The assumption of negligible population of the excited electronic state is valid so long as the detuning is sufficiently large, i.e.
\begin{equation}
\label{DeltaCondition}
 \tilde{g}_{b}\eta, h_b , \tilde{g}_c \ll \lvert\Delta\rvert.
 \end{equation}
We have also neglected the motional recoil due to spontaneous emission. This is justified when the spontaneous emission is isotropic and the system is deep in the Lamb-Dicke regime, so that
 \begin{equation}
\label{etaCondition}
\eta^2\Gamma \lesssim \tilde{g}_{b}\eta, h_b , \tilde{g}_c.
 \end{equation}
The Born-Markov assumption requires the memory time of the electronic degrees of freedom to be much shorter than the characteristic time scales of the effective evolution, which implies
\begin{equation}
\label{BornMarkovCondition}
k,  \kappa_{b/c},\delta E\ll \Gamma,
\end{equation}
where $\delta E$ represents any energy shift appearing in $\delta H_{abc}$. Finally, in order to put the master equation \eqref{effMEdoubleMain} into Lindblad form, we must perform a rotating-wave approximation, valid when
 \begin{equation}
  \label{rwaCondition}
k,  \kappa_{b/c},\delta E \ll \nu,
\end{equation}
which corresponds to the definition of the sideband-resolved regime for this system.

\subsection{Phonon dynamics}

\begin{table}[b]
\begin{tabular}{lcr} \toprule \\
Parameter  & Symbol & Value \\ \colrule
Hot temperature & $T_h$ & 5800~K \\
Room temperature & $T_r$ & 300~K \\
Trap frequency & $\nu/2\pi$ & 5~MHz\\
Lamb-Dicke parameter & $\eta$ & 0.041\\
Carrier frequency & $\varepsilon/2\pi$ & 810~THz\\
Spontaneous emission rate & $\Gamma/2\pi$ & 20~MHz \\
Trap heating rate & $\lambda$ & 10 quanta/s \\
Cavity misalignment & $d_b = d_c$ & 10~nm\\
\botrule
\end{tabular}
\caption{Table of parameters used in numerical calculations. \label{paramTab}}
\end{table}

In this subsection we characterise the performance of the refrigerator in terms of the steady-state phonon occupation $n_a(\infty)$, focusing specifically on cooling using sunlight as an energy source. We take representative parameters pertaining to \Yb (listed in Table~\ref{paramTab}). This species is a good choice due to its low mass and correspondingly small photon recoil, in addition to the existence of a closed dipole-allowed cooling transition. However, one could equally well consider other species of ion or neutral atom. 

From here on we set $g_{b/c} = g$ and $\kappa_{b/c} = \kappa$ for simplicity. We compute $n_a(\infty) = \Tr[a^\d a\rho_\infty]$ by solving $\dd\rho_\infty/\dd t = \LL\rho_\infty = 0$, with $\LL$ defined by the RHS of Eq.~\eqref{effMEdoubleMain}. We use a truncated Hilbert space with 71 phonon states and 4 states per cavity mode. Such a small Hilbert space dimension for the cavity modes is justified since the mean number of cavity photons in the steady state is $n_b(\infty) \approx 10^{-3}$ and $n_c(\infty) < 10^{-4}$ in cavities $b$ and $c$, respectively, for all parameters considered. We have checked that decreasing the Hilbert space dimension leads to negligible changes in the results.

Our predictions for $n_a(\infty)$ are shown in Fig.~\ref{doubleColourPlots}. We observe that sunlight at $T_h = 5800~$K is sufficient to drive the phonon almost to its ground state, so long as the effective coupling constant $k$ is sufficiently large. In Fig.~\ref{doubleColourPlots}(a) we show that, in the regime of effective cooling, the steady-state phonon occupation is reduced by increasing $\kappa$ for fixed $g$. Nevertheless, $\kappa$ must remain smaller than $\nu$ for the system to remain in the sideband-resolved regime (Eq.~\eqref{rwaCondition}), which represents a key factor limiting the achievable steady-state phonon occupation.  In Fig.~\ref{doubleColourPlots}(b) we demonstrate that increasing $\Delta$ for fixed $k$ can improve performance by suppressing incoherent effects associated with spontaneous emission.

In the limit of $\lvert\Delta\rvert \gg \Gamma$ and $k \ll \kappa$, we can give a rough analytical estimate of the relaxation time. In this regime we can derive an effective evolution equation for the motional degrees of freedom by tracing over the cavity modes, as shown in Appendix~\ref{appendixMotionalDynamics}. This approximate equation of motion can be solved to give the phonon population as a function of time:
\begin{equation}
\label{phononEvolutionApprox}
n_a(t) = n_\infty +\ee^{-\gamma t}  \left (n_0 - n_\infty\right ),
\end{equation}
where  $n_\infty = \lambda/\gamma$ is the steady-state phonon number, $n_0 = n_a(0)$ is the initial population, and the relaxation rate is $\gamma =k^2\bar{n}_b/\kappa$. 

\subsection{Collective coupling enhancement in many-ion systems}

In this subsection we generalise to the scenario where multiple atoms are trapped inside the cavities. We focus in particular on ion-trap systems, where the Coulomb interaction couples the motion of the different ions. The normal vibrational modes of the system are thus small collective oscillations about the mechanical equilibrium. We now show that if $N$ ions of the same species are placed inside the crossed-cavity refrigerator, an $N$-fold enhancement of the coupling between photons and phonons can be obtained.

The free Hamiltonian of the system is 
\begin{equation}
\label{H0multi}
H_2 = \nu a^\d a + \omega_b b^\d b + \omega_c c^\d c + \sum_{j=1}^N \varepsilon_j \sigma^+_j \sigma^-_j.
\end{equation}
Here, $a^\d$ creates a phonon of a normal mode with frequency $\nu/2\pi$, $\sigma^-_j$ is the atomic lowering operator for atom $j$ and we have allowed for variations of the electronic transition frequencies $\varepsilon_j$, due to inhomogeneous magnetic fields, for example. All other modes and electronic states are assumed to be off-resonant. The interaction Hamiltonian in the LDA and RWA reads as
\begin{align}
\label{interactionMulti}
V_2 = & \: \sum_{j=1}^N\left \lbrace \tilde{g}_b(\rr_j) \eta_j\left( a + a^\d \right) \left (b\sigma^+_j+ b^\d\sigma^-_j \right ) \right .\notag \\ &\left . + \;\tilde{g}_c(\rr_j) \left (c\sigma^+_j+ c^\d\sigma^-_j \right )  + h_b(\rr_j)\left (b\sigma^+_j+ b^\d\sigma^-_j \right )\right \rbrace,
\end{align}
where $\tilde{g}_{b/c}(\rr_j) = g_{b/c}(\rr_j)\cos\delta_{b/c}(\rr_j)$, $h_b(\rr_j) = g_b(\rr_j)\sin\delta_b(\rr_j)$, with $g_{b/c}(\rr_j)$ the cavity coupling constants for the ion with equilibrium position $\rr_j$, while $\eta_j$ are the Lamb-Dicke parameters and $\delta_{b/c}(\rr_j) = d_{b/c}(\rr_j)\omega_{b/c}/c_0$ are the dimensionless misalignments, where $d_{b}(\rr_j)$ ($d_{c}(\rr_j)$) is the distance in the $x$ direction ($y$ direction) between $\rr_j$ and the field node of cavity $b$ (anti-node of cavity $c$).

We assume again that the cavities are detuned from the electronic transition frequencies, $\omega_c = \varepsilon_j + \Delta_j = \omega_b+\nu$, with $\varepsilon_j \gg \lvert\Delta\rvert_j \gg g_{b/c}(\rr_j)$. After adiabatically eliminating the electronic excited states according to the procedure in Appendix \ref{appendixAdiabaticElimination}, we find an effective interaction of the form
\begin{equation}
\label{effInteractionMulti}
V_\mathrm{eff} = k_\mathrm{col} \left (abc^\d + a^\d b^\d c \right ).
\end{equation}
In the limit $\lvert\Delta_j\rvert \gg \Gamma$, the collective coupling constant is found to be
\begin{equation}
\label{effCouplingConstantMulti}
k_\mathrm{col} = \sum_{j=1}^N \frac{\tilde{g}_b(\rr_j)\tilde{g}_c(\rr_j)\eta_j}{\Delta_j}.
\end{equation}

The effective collective coupling can be either enhanced or suppressed compared to the single-particle case, depending on the symmetry of the normal mode in question. For example, let us take $N=2$ and assume that $\tilde{g}_{b/c}(\rr_1) = \tilde{g}_{b/c}(\rr_2) = \tilde{g}_{b/c}$ and $\Delta_1 = \Delta_2 = \Delta$. For the stretch mode, with $\eta_1 = -\eta_2$, we find that $k_\mathrm{col}=0$. On the other hand, for the centre-of-mass mode, with $\eta_1 = \eta_2 = \eta$, we find a two-fold enhancement of the coupling, i.e.\ $k_\mathrm{col} = 2 k$, where $k = \tilde{g}_b\tilde{g}_c\eta/\Delta$ is the single-ion effective coupling. In general, the centre-of-mass oscillations experience enhanced collective coupling, since for this vibrational mode $\eta_j = \eta$ is constant. Therefore, it is possible to improve the refrigerator's performance when cooling the centre-of-mass mode by incorporating many ions inside the cavities. 

Note that the heating rate may also increase with the number of ions. However, the estimate given by Eq.~\eqref{phononEvolutionApprox} indicates that the collective coupling enhancement still leads to an improvement in the achievable temperature and relaxation time so long as the heating rate increases slower than $N^2$.

\section{Conclusion	} 
\label{conclusion}

To summarise, we have analysed the possibility of creating an autonomous thermal machine (ATM) in the setting of cavity quantum electrodynamics (CQED). In particular, we have shown that it is possible to construct a refrigerator which cools the motion of a trapped atom using sunlight. This would constitute perhaps the first example of a quantum absorption chiller which can achieve technologically useful temperatures.

In principle, such a refrigerator powered by sunlight costs nothing to operate in daylight hours (under clement weather conditions). This is clearly an improvement on power-hungry and thermodynamically inefficient cooling lasers. In practice however, we find that the technical difficulty of stabilising the cavity frequencies makes the use of a laser, or similar frequency reference system, unavoidable with present technology. 

Commercially available laser systems enjoy stability, reliability and flexibility, spanning a range of optical and ultra-violet frequencies. These properties are unmatched by optical cavities currently available in CQED experiments. Therefore, absorption chillers of the kind we envisage are unlikely to supersede laser-driven cooling schemes in the near future. Nevertheless, our results demonstrate that if the intrinsic stability of optical resonators can be improved, ATMs could in principle play a useful role in quantum state preparation.

A more immediately relevant feature of our proposal is that it offers a versatile experimental platform to study the physics of ATMs. One advantage of our scheme is that the frequencies of --- and the couplings between --- different degrees of freedom are tunable by means of external control fields, or by modifying the cavity lengths mechanically. We have also shown that the coupling can be enhanced $N$-fold when $N$ ions of the same species are incorporated inside the refrigerator. Note that experiments demonstrating collective cavity coupling enhancement in trapped-ion systems have already achieved $N\sim 500$ \cite{Herskind2009nat,Albert2012pra}, implying that very large three-body interaction energies could be obtained. In such a regime, our simple local dissipation model is invalid, and delocalised dissipation effects should become important \cite{Correa2013pre}.

Another attractive feature of the CQED setting is the wide variety of measurements that are available. For example, a different species of ion placed inside the trap could be used to make non-demolition measurements of steady-state properties, such as the phonon number \cite{Bermudez2013prl}. One could also continuously and non-destructively monitor the state of the cavity fields using appropriately placed photodetectors. This would yield direct information on the rate of energy dissipation into the environment, as well as opening up a new potential arena for exploring quantum thermodynamics with measurement feedback \cite{Brandner2015njp,Vidrighin2016prl}.  Non-thermal or non-Markovian reservoirs could be engineered by modifying the spectrum or statistics of the radiation incident on the cavity \cite{Murch2012prl}, or by tickling the ion-trap electrodes with suitably filtered electrical noise \cite{Myatt2000nat}. We also note that analogous set-ups using different kinds of quantum emitters such as superconducting qubits, quantum dots or diamond color centers may be envisaged. 

On a conceptual level, our study provides a number of further insights. We found a simple and intuitive interpretation of quantum absorption refrigerators in terms of sideband transitions. The present context of atomic cooling makes the connection between these two concepts manifest, however this link is in fact completely general. Any three-body absorption refrigerator can be understood in terms of a red sideband transition, which is excited by the absorption of quanta from a thermally populated auxiliary system, itself connected to a hot reservoir. The role of the second, colder reservoir is to then quickly reset the state of the refrigerator by a transition at the carrier frequency (that is, at frequency $\varepsilon$), leaving the target system with one less quantum of energy. Thus, sideband transitions provide an alternative framework for understanding quantum ATMs which is complementary to the standard picture in terms of virtual qubits and temperatures.

We also found that, just as in laser sideband cooling, the existence of blue sideband transitions limit the thermal dissipation rates and the three-body interaction strength to be less than the frequency of the target subsystem, i.e. one must be in the sideband-resolved regime. This constraint is rather general, because blue sideband transitions (or more generally, off-resonant transitions) exist for any absorption refrigerator governed by an interaction Hamiltonian composed of a single product of Hermitian operators (rather than a sum of such products). We thus expect a similar sideband-resolved condition to generically constrain the achievable power and other relevant thermodynamic quantities describing these machines. 

As we have shown, an absorption refrigerator transferring energy from motional degrees of freedom to optical photons can achieve remarkably low temperatures in principle. This results from  the huge separation between vibrational and optical frequencies, leading to extremely small virtual temperatures (see Eq.~\eqref{virtualTempDouble}). The same principle underlies other recent proposals to build thermal machines using cavity optomechanical systems \cite{Mari2012prl, Zhang2014prl}. However, this separation of frequency scales also implies an instability of the system to relatively small fluctuations or drifts of the optical frequencies. This suggests that a practical operating regime for truly autonomous quantum thermal machines will be such that the natural frequencies of the constituent subsystems are commensurate with each other. 

In the present example, the effects of frequency drift can be overcome by weakly driving the cavity with a laser. This indicates that the truly essential resource for cooling in the quantum regime is a stable frequency reference, or equivalently an \textit{accurate clock}. Given such a frequency reference, we have shown that chaotic thermal energy suffices to cool the system almost to its ground state. If one adopts the view that the heat energy driving the absorption refrigerator is a free resource, the performance of the machine is then ultimately limited by the efficiency of the frequency reference or clock. A natural question thus arises regarding the fundamental thermodynamic limitations on clocks, accounting for the energy required to measure the clock \cite{Micadei2013pre} and any effect of correlations between the clock and the thermal machine \cite{Malabarba2015njp, MischaInPrep}. This intriguing problem will be tackled in future publications.

\section{Acknowledgements}

We gratefully acknowledge edifying conversations with Nikolai Kiesel, Alexander Kubanek, Joseph Randall, Johannes Ro{\ss}nagel, Kilian Singer, Raam Uzdin and Mihai Vidrighin. MTM and MPW were financed by EPSRC. MH acknowledges support from the Swiss National Science Foundation (AMBIZIONE PZ00P2$\_$161351) and the MINECO  project FIS2013-40627-P, with the support of FEDER funds, and by the Generalitat de Catalunya CIRIT, project 2014-SGR-966. JP was supported by Minister{\'i}o de Econom{\'i}a y Competitividad  Project No.~FIS2015-69512-R and the Fundaci{\'o}n S{\'e}neca Project No.~19882/GERM/15. MBP was supported by an Alexander von Humboldt Professorship, the ERC Synergy grant BioQ and the EU STREP EQUAM.

\bibliographystyle{unsrt}
\bibliography{cavityFridge}
  
\appendix
\clearpage
\begin{widetext}

\section{Projection onto the electronic ground state}
\label{appendixAdiabaticElimination}

In this appendix we explain how to perturbatively eliminate the excited electronic state from the equations of motion. Our approach closely follows the treatment of Refs.~\cite{Breuer2007book,Rivas2010njp}; similar analyses can also be found in Refs.~\cite{QuantumNoise,VanKampen}. Although much of the following is textbook material, an additional complication is introduced by the dissipative coupling between the motional and electronic degrees of freedom due to momentum recoil from spontaneously  emitted photons. We therefore present each step of the derivation in detail. 

Our starting point is the master equation describing the motional, cavity, and electronic degrees of freedom, which can be written as
\begin{equation}
\label{masterEquationGeneral}
\dt{\chi} = \mathcal{L}\chi.
\end{equation}
The objective is to trace over the electronic variables, leaving an effective master equation describing the density matrix $\rho = \Tr_\sigma[\chi]$ of the remaining degrees of freedom. We write the local Hamiltonian as $H_2 =  H_{abc} + H_\sigma $, where
\begin{align}
\label{Habc}
H_{abc} & = \nu a^\d a + \omega_b b^\d b + \omega_c c^\d c,\\
\label{He}
H_\sigma & =  \varepsilon\sigma^+ \sigma^-.
\end{align}
The interaction in the Lamb-Dicke and rotating-wave approximations is
\begin{equation}
V_2 =  \tilde{g}_b \eta \left (a + a^\d \right ) \left ( b  \sigma^+ + b^\d \sigma^- \right ) + h_b \left ( b  \sigma^+ + b^\d \sigma^- \right ) + \tilde{g}_c  \left ( c\sigma^+ + c^\d\sigma^- \right ). 
\end{equation}
We use the following symbols to denote commutation superoperators:
\begin{equation}
\label{Hsuperop}
\HH_\sigma\chi = -\ii[H_\sigma,\chi], \qquad \HH_{abc}\chi = -\ii[H_{abc},\chi], \qquad \VV \chi = -\ii[V_2,\chi].
\end{equation}
The Liouvillian can be decomposed into three contributions as $\LL = \LL_0 + \LL_{1} + \VV$, with $\LL_0 = \HH_{abc} + \HH_\sigma + \LL_\sigma$ and $\LL_1 = \LL_a + \LL_b + \LL_c$. In the following we set $\bar{n}_\sigma = 0$ in $\LL_\sigma$ (see Eq.~\eqref{Lsig}), which is an excellent approximation for optical frequencies at room temperature.

We now introduce a projector onto the electronic ground state $\PP\chi = \Tr_\sigma[\chi]\otimes\ket{\downarrow}\bra{\downarrow}$, and its orthogonal complement $\QQ = 1 - \PP$. We assume that the electron is in its ground state at $t=0$, and uncorrelated with $S$, which implies that $\QQ\chi(0) = 0$. 
Now we move to a dissipation picture defined by
\begin{equation}
\label{dissipationPictureState}
\tilde{\chi}(t) = \ee^{-\LL_0 t} \chi(t),
\end{equation}
\begin{equation}
\label{dissPictureOps}
\VV(t) = \ee^{-\LL_0 t} \VV\ee^{\LL_0t},\qquad \LL_1(t) = \ee^{-\LL_0 t} \LL_1\ee^{\LL_0t}.
\end{equation}
It is important to note that the action of the superoperators $\LL_0$, $\LL_1$ and $\VV$ is not associative: they are defined to operate on everything that appears to their right. We also note the useful identities
\begin{align}
\label{L0P}
\LL_0\PP & = \HH_{abc}\PP, \\
\label{PVP}
\PP\VV(t)\PP & = 0, \\
\label{PLP}
\PP\LL_1(t)\PP & = \LL_1(t)\PP, \\
\label{PLVP}
\PP\LL_1(t)\VV(t')\PP & = 0.
\end{align}
These expressions can be proved by considering their action on a general quantum state.

In terms of their typical eigenvalues, we have that $\LL_0 \gg \LL_1,\VV$. This allows us to perturbatively eliminate the irrelevant part of the density operator $\QQ \chi(t)$. In order to do this, we write the master equation in the dissipation picture as
\begin{equation}
\label{dissPictureME}
\dt{\tilde{\chi}} = \LL_1(t)\tilde{\chi}(t) + \VV(t)\tilde{\chi}(t),
\end{equation}
and insert the identity $1 = \PP + \QQ$ on both sides, finding
\begin{align}
\label{P-EqOfMotion}
\dt{\PP\tilde{\chi}} &= \LL_1(t)\PP\tilde{\chi}(t) + \PP\LL_1(t)\QQ\tilde{\chi}(t) + \PP\VV(t)\QQ\tilde{\chi}(t), \\
\label{Q-EqOfMotion}
\dt{\QQ\tilde{\chi}} & = \QQ\LL_1(t)\QQ\tilde{\chi}(t) + \QQ\VV(t)\QQ\tilde{\chi}(t) + \VV(t)\PP\tilde{\chi}(t),
\end{align}
where Eqs.~\eqref{PVP} and \eqref{PLP} have been used. Eq.~\eqref{Q-EqOfMotion} can be formally solved by introducing the propagator 
\begin{equation}
\label{propagator}
\mathcal{G}(t,t') =  \mathbf{T}\exp \left ( \int_{t'}^{t} \dd s\; \QQ\left [  \LL_1(s) + \VV(s) \right ]\QQ \right ),
\end{equation}
where the symbol $\mathbf{T}$ denotes the usual time ordering. The solution for $\QQ\tilde{\chi}$ is
\begin{equation}
\label{Qsoln}
\QQ\tilde{\chi}(t) = \mathcal{G}(t,0)\QQ\chi(0) + \int_0^t \dd t'\; \mathcal{G}(t,t') \VV(t')\PP\tilde{\chi}(t'),
\end{equation} 
and the first term on the RHS vanishes for our choice of initial conditions. Substituting the solution Eq.~\eqref{Qsoln} into Eq.~\eqref{P-EqOfMotion}, we obtain an exact evolution equation for $\PP\tilde{\chi}$:
\begin{equation}
\label{P-closedFormalEoM}
\dt{\PP\tilde{\chi}} = \LL_1(t)\PP\tilde{\chi}(t) +  \int_0^t \dd t'\;\PP\left [\LL_1(t) + \VV(t)\right ] \mathcal{G}(t,t') \VV(t')\PP\tilde{\chi}(t').
\end{equation}

At this stage we approximate Eq.~\eqref{P-closedFormalEoM} by expanding the RHS up to second order in the small quantities $\LL_1$ and $\VV$, which yields
\begin{equation}
\label{PsolutionPerturbative}
\dt{\PP\tilde{\chi}} = \LL_1(t)\PP\tilde{\chi}(t) +  \int_0^t \dd t'\;\PP \VV(t) \VV(t')\PP\tilde{\chi}(t'),
\end{equation}
where we have used Eq.~\eqref{PLVP}. We also note that
\begin{equation}
\label{LsigmaPerturbative}
\LL_\sigma = \Gamma \DD[\sigma^-] + O(\eta^2\Gamma),
\end{equation}
assuming that the angular emission distribution is symmetric, $\Pi(u) = \Pi(-u)$, which holds true for spontaneous emission in an isotropic environment. We assume that $\eta^2 \Gamma$ is on the same order as $\LL_1$ and $\VV$, which is the case deep in the Lamb-Dicke regime $\eta\ll 1$. To second order in small quantities, it is therefore sufficient to retain only the leading-order contribution $\LL_\sigma \approx \Gamma\DD[\sigma^-]$ in evaluating the second term on the RHS of Eq.~\eqref{PsolutionPerturbative}.

We now invoke the Markov approximation by extending the lower integration limit to $t'=-\infty$ and making the replacement $\tilde{\chi}(t') \to \tilde{\chi}(t)$ . These steps are justified because the memory kernel $\PP \VV(t)\VV(t')\PP$ decays rapidly to zero. In particular, this decay is approximately exponential in time with decay constant $2/\Gamma$, which is much shorter than the characteristic time scales of the reduced system dynamics. After a change of variables to $s = t-t'$, we obtain the Markovian master equation
\begin{equation}
\label{markovMasterEquation}
\dt{\PP \tilde{\chi}} = \LL_1(t)\PP\tilde{\chi}(t) + \int_0^\infty \dd s\; \PP \VV(t) \VV(t-s) \PP \tilde{\chi}(t).
\end{equation}
To evaluate this expression explicitly, is convenient to introduce a decomposition of the interaction Hamiltonian as 
\begin{equation}
\label{V1decomposition}
V = \sum_{\alpha} \sum_{\Omega} L_\alpha(\Omega)\otimes R_\alpha,
\end{equation}
where the sum over $\Omega$ in Eq.~\eqref{V1decomposition} runs over all Bohr frequencies of $H_S$, while the $L_\alpha(\Omega)$ are lowering operators for these frequencies, i.e.
\begin{equation}
\label{eigenoperators}
[H_S,L_\alpha(\Omega)] = -\Omega L_\alpha(\Omega),
\end{equation}
and the operators $R_\alpha$ act only on the electronic degrees of freedom. Substituting Eq.~\eqref{V1decomposition} into Eq.~\eqref{markovMasterEquation} and tracing over the electronic variables, we obtain
\begin{equation}
\label{redfieldEqnExplicit}
\Tr_\sigma \left [ \int_0^\infty \dd s\; \PP \VV(t) \VV(t-s) \PP \tilde{\chi}(t)\right ] = \sum_{\alpha,\beta} \sum_{\Omega,\Omega'} G_{\alpha\beta}(\Omega) \ee^{\ii(\Omega'-\Omega)t}\left [ L_{\alpha}(\Omega)\tilde{\rho}(t) L^\d_\beta(\Omega') - L_\beta^\d(\Omega') L_\alpha(\Omega)\tilde{\rho}(t) \right ] + \mathrm{h.c.},
\end{equation}
where $\tilde{\rho}(t) = \ee^{\ii H_{abc}t} \Tr_\sigma[\chi(t)] \ee^{-\ii H_{abc}t}$, and we defined the spectral correlation matrix
\begin{equation}
\label{selfEnergy}
G_{\alpha\beta}(\Omega) = \int_0^\infty \dd t\; \ee^{\ii\Omega t} \bra{\downarrow} R_\beta^\d(t) R_\alpha \ket{\downarrow}.
\end{equation}
Here, $R^\d_\beta(t) = \ee^{\mathcal{K}_\sigma t}[R^\d_\beta],$ where $\mathcal{K}_{\sigma} = \HH_\sigma + \Gamma\DD[\sigma^-]$ and its adjoint $\mathcal{K}^\dagger_\sigma$ is defined by $\Tr\{\mathcal{K}_\sigma^{\d}[A] B\} = \Tr\{A\mathcal{K}_\sigma[B] \}$. The correlation functions can be decomposed as 
\begin{align}
\label{gamma}
\gamma_{\alpha\beta}(\Omega	) & = G_{\alpha\beta}(\Omega) + G_{\beta\alpha}^*(\Omega),  \\
\label{shift}
S_{\alpha\beta}(\Omega) & = \frac{1}{2\ii} \left (  G_{\alpha\beta}(\Omega) - G_{\beta\alpha}^*(\Omega)  \right ).
\end{align}
The final step is the rotating wave approximation, in which rapidly oscillating contributions with $\Omega \neq \Omega'$ are neglected. Transforming back to the Schr\"{o}dinger picture yields the Lindblad master equation
\begin{equation}
\label{finalMErwa}
\dt{\rho} = -\ii[H_{abc} + H_L,\rho] + \LL_1\rho + \sum_{\alpha,\beta}\sum_{\Omega}\gamma_{\alpha\beta}(\Omega) \left (L_\alpha(\Omega)\rho L^\d_{\beta}(\Omega) - \frac{1}{2}\{L^\d_{\beta}(\Omega)L_\alpha(\Omega),\rho\} \right ),
\end{equation}
with 
\begin{equation}
\label{lambShift}
H_L = \sum_{\alpha,\beta} \sum_{\Omega} S_{\alpha\beta}(\Omega) L_\beta^\d(\Omega) L_\alpha(\Omega).
\end{equation}
The master equation is then placed into Lindblad form by diagonalising the matrices $\gamma_{\alpha\beta}(\Omega)$ \cite{Breuer2007book}.

The non-vanishing components of the spectral correlation matrix are proportional to
\begin{equation}
\label{electronicCorrFunc}
\int_0^\infty\dd t\; \ee^{\ii\Omega t} \bra{\downarrow}\sigma^-(t)\sigma^+ \ket{\downarrow} = \frac{2\Gamma + 4\ii(\Omega-\varepsilon)}{\Gamma^2 + 4(\Omega-\varepsilon)^2}.
\end{equation}
The Lamb-shift Hamiltonian is given as $H_L = \delta H_{abc} + V_\mathrm{eff}$, where
\begin{align}
\label{deltaHabc}
\delta H_{abc} = & \left [\frac{\tilde{g}_b^2\eta^2\Delta}{\Gamma^2/4 + \Delta^2} + \frac{\tilde{g}_b^2\eta^2(\Delta-2\nu)}{\Gamma^2/4 + (\Delta- 2\nu)^2}\right ] a^\d a b^\d b + \left [\frac{\tilde{g}_b^2\eta^2(\Delta-2\nu)}{\Gamma^2/4 + (\Delta- 2\nu)^2} + \frac{h_b^2(\Delta-\nu)}{\Gamma^2/4+(\Delta-\nu)^2}\right ] b^\d b \notag \\
& + \: \frac{\tilde{g}_c^2\Delta}{\Gamma^2/4 + \Delta^2} c^\d c,
\end{align}
\begin{equation}
\label{VeffFull}
V_\mathrm{eff} = \frac{\tilde{g}_b \tilde{g}_c \eta \Delta}{\Gamma^2/4 + \Delta^2} \left ( a b c^\d + a^\d b^\d c\right ).
\end{equation}
We also find the following incoherent contributions associated with spontaneous emission from the excited state:
\begin{align}
\label{Lse}
\LL_\mathrm{se} = & \frac{\Gamma\sqrt{\tilde{g}_b^2\eta^2 + \tilde{g}_c^2}}{\Gamma^2/4 + \Delta^2} \DD[\tilde{g}_b\eta ab + \tilde{g}_c c] + \frac{\tilde{g}_b^2\eta^2 \Gamma}{\Gamma^2/4 + (\Delta -  2\nu)^2} \DD[a^\d b] + \frac{h_b^2\Gamma}{\Gamma^2/4 + (\Delta-\nu)^2}\DD[b].
\end{align}
We see that for $\lvert\Delta\rvert > \Gamma$, the contributions from spontaneous emission $\LL_\mathrm{se}$ are suppressed by a factor of order $\Gamma/\Delta$ with respect to the coherent contribution of $H_L$. In the limit $\lvert\Delta\rvert \gg \Gamma,\nu$, and after dropping terms of second order in the small quantities $\eta$ and $\delta_{b/c}$, we obtain simply
\begin{equation}
\label{HLsDoubleLimit}
\delta H_{abc} \approx \frac{\tilde{g}_c^2}{\Delta} c^\d c,
\end{equation}
\begin{equation}
\label{VeffLimit}
V_\mathrm{eff} \approx \frac{\tilde{g}_b\tilde{g}_c\eta}{\Delta} \left ( a b c^\d + a^\d b^\d c\right ),
\end{equation}
\begin{equation}
\label{LseLimit}
\LL_\mathrm{se} \approx  \frac{\Gamma\sqrt{\tilde{g}_b^2\eta^2 + \tilde{g}_c^2}}{\Delta^2} \DD[\tilde{g}_b \eta ab + \tilde{g}_c c].
\end{equation}

Finally, we comment on the case where multiple trapped ions are placed inside the refrigerator system. Assuming that the environmental fluctuations seen by each ion are uncorrelated, then the derivation is essentially unchanged. One finds separate, additive contributions from each ion of the same form as Eqs.~\eqref{deltaHabc}, \eqref{VeffFull} and \eqref{Lse}. The full time evolution generator is given by a sum over these contributions, in addition to the free Hamiltonian $H_{abc}$. In particular, summing over the contributions corresponding to Eq.~\eqref{VeffFull} leads directly to Eq.~\eqref{interactionMulti}, in the limit $\lvert\Delta\rvert \gg \Gamma$.

\section{Approximate motional dynamics}
\label{appendixMotionalDynamics}

In this appendix we derive a simple approximate model for the dynamics of the atomic motion that is easy to solve analytically. We start from the master equation~\eqref{finalMErwa}, and aim to derive an evolution equation for the motional density matrix $\rho_a(t) = \Tr_{bc}[\rho(t)]$ obtained by tracing out the cavity modes. The procedure is similar to that of Appendix~\eqref{appendixAdiabaticElimination}. For simplicity, we work in the limit of large detunings, so that $\LL_\mathrm{se}$ can be neglected. We also ignore higher-order corrections contributed by $\delta H_{abc}$.

The derivation proceeds as follows. We define a projection operator $\PP\rho = \Tr_{bc}[\rho]\otimes \rho_{b}\otimes\rho_c$, where $\rho_{b/c}$ are thermal states of modes $b$ and $c$ at temperatures $T_h$ and $T_r$, respectively. We also assume that the initial quantum state factorises such that $\QQ\rho(0) = (1-\PP)\rho(0) = 0$. We split the Liouvillian into parts as $\LL = \LL_0 + \LL_a + \VV_\mathrm{eff}$, where $\LL_0 = \HH_{abc} + \LL_b + \LL_c$, while $\VV_\mathrm{eff}$ is the commutation superoperator generated by the effective interaction~\eqref{VeffLimit}. Moving to a dissipation picture generated by $\LL_0$,
\begin{equation}
\label{dissipationPictureState2}
\tilde{\rho}(t) = \ee^{-\LL_0 t} \rho(t),
\end{equation}
\begin{equation}
\label{dissPictureOps2}
\VV_\mathrm{eff}(t) = \ee^{-\LL_0 t} \VV_\mathrm{eff}\ee^{\LL_0t},\qquad \LL_a(t) = \ee^{-\LL_0 t} \LL_a\ee^{\LL_0t}.
\end{equation}
one readily verifies the following properties
\begin{align}
\label{L0P2}
\LL_0\PP = \HH_a\PP, \\
\label{PVP2}
\PP\VV_\mathrm{eff}(t)\PP & = 0, \\
\label{PLP2}
[\PP,\LL_a(t)] & = 0, \\
\label{LVcommute}
[\LL_a(t),\VV_\mathrm{eff}(t')] & = 0.
\end{align}
where $\HH_a$ is the commutation superoperator associated with $H_a = H_{abc} - H_{bc} = \nu a^\d a$. 

We assume that $\lambda \ll \kappa_{b/c}$ and $k\ll \kappa_{b/c}$, meaning that the term $\LL_0$ dominates and the other contributions $\LL_a$ and $\VV_\mathrm{eff}$ can be accounted for perturbatively. Using the projection operators and the properties~\eqref{PVP2}, \eqref{PLP2} and \eqref{LVcommute}, we derive a closed equation of motion for the relevant part of the density matrix:
\begin{equation}
\label{markovMasterEquation2}
\dt{\PP \tilde{\rho}} = \LL_a(t)\PP\tilde{\rho}(t) + \int_0^\infty \dd s\; \PP \VV_\mathrm{eff}(t) \VV_\mathrm{eff}(t-s) \PP \tilde{\rho}(t).
\end{equation}
Here we have made a Born-Markov approximation, which is justified in the limit $k\ll \kappa_{b/c}$. The subsequent formal manipulations proceed exactly as in Appendix~\ref{appendixAdiabaticElimination}, in particular the part following Eq.~\eqref{markovMasterEquation}. The relevant elements of the spectral correlation matrix are
\begin{align}
\label{spectCorrMatrix2}
\int_0^\infty\dd t \;\ee^{\ii\nu t} \langle b^\d(t)b(0) c(t) c^\d(0)\rangle & \approx \frac{\bar{n}_b}{\kappa_b+\kappa_c}\notag \\
\int_0^\infty\dd t\; \ee^{-\ii\nu t} \langle b(t)b^\d(0) c^\d(t) c(0)\rangle & \approx 0.
\end{align}
Here, the time evolution of the operators is given by $b(t) = \ee^{\LL_0^\d t} b$ and $c(t) = \ee^{\LL_0^\d t} c$, while the angle brackets denote an average with respect to a thermal product state of the cavities, i.e. $\rho_b\otimes \rho_c$, and we used the fact that $\bar{n}_c\approx 0$. The final master equation thus takes the form
\begin{equation}
\label{motionalME}
\dt{\rho_a} = -\ii [ \nu a^\d a, \rho_a] + (\lambda +\gamma)\DD[a]\rho_a + \lambda \DD[a^\d]\rho_a,
\end{equation}
where $\gamma = 2k^2\bar{n}_b/(\kappa_b+\kappa_c)$, and we used the fact that $\bar{n}_a^{-1}\approx 0$.

\end{widetext}

\end{document}